\documentclass[twocolumn,aps,superscriptaddress,showpacs,floatfix,prc]{revtex4}
\usepackage{mathrsfs}
\usepackage{amssymb}
\usepackage{amsmath}
\usepackage{graphicx}
\usepackage[normalem]{ulem}
\usepackage[dvips]{color}
\usepackage{bm}
\usepackage{longtable}
\usepackage{slashed}
\usepackage{epstopdf}

\usepackage{times}

\renewcommand\sout{\bgroup \color{red} \ULdepth=-.5ex \ULset}

\renewcommand{\v}[1]{\textbf{#1}}
\renewcommand{\rm}[1]{\textrm{#1}}
\renewcommand{\d}{\mathrm{d}}

\begin{document}

\title{Neutron matter within QCD sum rules}

\author{Bao-Jun Cai}
\affiliation{Department of Physics, Shanghai University, Shanghai
200444, China} \affiliation{School of Physics and Astronomy and
Shanghai Key Laboratory for Particle Physics and Cosmology, Shanghai
Jiao Tong University, Shanghai 200240, China}
\author{Lie-Wen Chen\footnote{%
Corresponding author: lwchen$@$sjtu.edu.cn}} \affiliation{School of
Physics and Astronomy and Shanghai Key Laboratory for Particle
Physics and Cosmology, Shanghai Jiao Tong University, Shanghai
200240, China} \affiliation{Center of Theoretical Nuclear Physics,
National Laboratory of Heavy Ion Accelerator, Lanzhou 730000, China}
\date{\today}

\begin{abstract}

Equation of state (EOS) of pure neutron matter (PNM)
is studied in QCD sum rules (QCDSR).
It is found that the QCDSR results on EOS of PNM are in good agreement
with predictions by current advanced microscopic many-body theories.
Moreover, the higher-order density terms in quark condensates are
shown to be important to describe the empirical EOS of PNM {in the density region} around and
above nuclear saturation density {although they play minor role at subsaturation densities.}
The chiral condensates in PNM are also studied, and our results
indicate that the higher-order density terms in quark condensates,
which are introduced to reasonably describe the empirical EOS of PNM at
suprasaturation densities, tend to hinder the appearance of chiral
symmetry restoration in PNM at high densities.
\end{abstract}

\maketitle

%\tableofcontents

\section{Introduction}

Equation of state (EOS) of cold pure neutron matter (PNM) is an
interesting and important problem at least from two aspects. On the
one hand, at sub-saturation even to very low densities, the PNM
composed of spin-down and -up neutrons with a large s-wave
scattering length shows several universal properties\,\cite{Tan08}
such as the simplicity of its EOS characterized by a few universal
parameters\,\cite{Cai15a,Kru15,Kol16,ZhaNB17}; the high momentum
tail above the Fermi surface of the single nucleon momentum
distribution function in cold PNM is also found to be very similar
to that in ultra-cold atomic Fermi gases\,\cite{Hen14} although the
magnitude of the density for the two systems differs by about $25$
orders\,\cite{Hen15}. Thus, the cold PNM at low densities
provides a perfect testing bed to explore novel ideas in the
unitary region\,\cite{Gio08,Blo08}, helping to find
deep physical principles behind these quantum many-body
systems\,\cite{Zwe12}. On the other hand, cold PNM at densities up
to 3-5$\rho_0$, with $\rho_0 \sim 0.16$ fm$^{-3}$ the nuclear saturation density, is
extremely important to the properties of neutron
stars\,\cite{Gle00,Lat04,Lat12,Oze16,EPJA}, such as the mass-radius
relation of a neutron star and its transport properties\,\cite{Hae07},
since the EOS of neutron star matter is very close to that of PNM.

Conventionally, since there lack direct experimental probes on the
PNM~\cite{Zha15}, people usually rely on phenomenological
models~\cite{Sto06,Ser86} to explore its properties.
However, due to the fact that the fitting scheme in these models is
usually implemented by a number of phenomenological parameters, the
microscopic origin of the uncertainties on the EOS of PNM are often averaged.
Consequently, any microscopic approaches to EOS of PNM, especially those
inheriting the quantum chromodynamics (QCD) spirit,
such as the effective field theories\,\cite{Tew13,Kru13,Kru13a,Dre14,Hag14,Heb15,Dri16} and
simulations\,\cite{Car15}, are appealing and exciting.

The QCD sum rules (QCDSR) method\,\cite{Shi79a} provides an important
non-perturbative QCD approach to explore the properties of nuclear matter (see, e.g., Refs.\,\cite{Dru90}).
Intuitively, when the QCD coupling constant is small at high
energies/small distances, the theory becomes asymptotically free,
guaranteeing the applicability of perturbative calculations. As the
energy scale decreases, the coupling constant of the theory becomes
large, perturbative methods break down eventually and
non-perturbative effects emerge. Among these effects, the most
important is the appearance of the quark/gluon condensates.
The basic idea of QCDSR for nuclear
matter calculations\,\cite{Dru90,Coh91,Fur92,Jin93,Dru04,Dru17,Jeo13,Coh95,Iof11} is to relate these condensates to the nucleon
self-energies using the operator product expansion (OPE) technique,
where information on the self-energies is introduced via nucleon-nucleon
correlation functions.
Within the QCDSR {method}, the exact information on the nucleon self-energies and
nuclear matter EOS can thus provide constraints on the in-medium quark
condensate, which is an order parameter of spontaneous
chiral symmetry breaking in QCD.
The QCDSR {method} is expected to work well at lower
densities/momenta where effects of the poorly-known high mass-dimensional
condensates as well as continuum effects are small enough.

In this work, we mainly focus on the properties of PNM
obtained with the QCDSR, and leave the detailed descriptions and
more physical issues about asymmetric nuclear matter to be reported
elsewhere\,\cite{Cai17x}. The EOS of PNM defined by the energy per
neutron can be obtained as\,\cite{Ber88,XuC11}
\begin{equation}\label{EOSPNM}
E_{\rm{n}}(\rho)=\frac{1}{\rho}
\int_0^{\rho}\d\rho\left[e_{\rm{n}}^{\ast}(\rho)+\Sigma_{\rm{V}}^{\rm{n}}(\rho,k_{\rm{F}}^{\rm{n}})-M\right],
\end{equation}
where we denote
$e_{\rm{n}}^{\ast}(\rho)=[(M+\Sigma_{\rm{S}}^{\rm{n}}(\rho,k_{\rm{F}}^{\rm{n}}))^2+k_{\rm{F}}^{\rm{n},2}]^{1/2}$
with $M$ the nucleon rest mass, and
$\Sigma_{\rm{S/V}}^{\rm{n}}(\rho,k_{\rm{F}}^{\rm{n}})$ is the
scalar/vector self-energy of a neutron in PNM at its Fermi surface $k_{\rm{F}}^{\rm{n}}=(3\pi^2\rho)^{1/3}$. This Lorentz
structure of the single neutron energy
$e_{\rm{n}}=e_{\rm{n}}^{\ast}+\Sigma_{\rm{V}}^{\rm{n}}$ is very
general owing to the translational/rotational/parity and
time-reversal invariance as well as the hermiticity in the rest
frame of neutron matter\,\cite{Ser86,Cai12}. The main motivation of
this work is to obtain the $E_{\rm{n}}(\rho)$ by
Eq.\,(\ref{EOSPNM}) with the density and momentum dependent self-energies, i.e., $\Sigma^{\rm{n}}_{\rm{S/V}}(\rho,|\v{k}|)$,
determined by the QCDSR.

Successes of QCDSR in nuclear matter calculations can be traced back to the
prediction on the large nucleon self-energies on GeV
scale\,\cite{Coh91}. And the present work is a natural generalization to
the study of PNM with QCDSR. As we shall see, the
results on EOS of PNM and quark condensates at low densities obtained via the
QCDSR are consistent with predictions by other state-of-the-art microscopic many
body theories, demonstrating that QCDSR can be applied to explore properties of {PNM} quantitatively.

Section \ref{Sec2} briefly introduces the QCDSR method. In section
\ref{Sec3}, the results on the $E_{\rm{n}}(\rho)$ from the QCDSR are
{presented}. Section \ref{Sec4} is
devoted to the study on the chiral condensates in PNM.
Section \ref{Sec5} is the summary of this work.

\section{A Brief Introduction to QCDSR}\label{Sec2}

As discussed in the introduction, the essential task of the QCDSR calculations for nuclear matter is
to relate, via OPE, the quark/gluon condensates with the nucleon
self-energies, and the latter are usually encapsulated in the
nucleon-nucleon correlation functions $\Pi_{\mu\nu}$ constructed by
quantum hadrodynamics\,\cite{Ser86}. The form of $\Pi_{\mu\nu}$ at
zero density (vacuum) is generally given by\,\cite{Coh95}
\begin{align}\hspace*{-0.2cm} \Pi_{\mu\nu}(q)\equiv&
i\int\d^4xe^{iqx}\langle0|\rm{T}\eta_{\mu}(x)\overline{\eta}_{\nu}(0)|0\rangle\notag\\
=&-\int\d
a_0\left[\frac{\rho_{\mu\nu}(a)}{q_0-a_0+i0^+}+\frac{\widetilde{\rho}_{\mu\nu}(a)}{q_0-a_0-i0^+}\right],\label{SP}
\end{align}
where $q$ is the momentum transfer and $a=(a_0,\v{q})$, $|0\rangle$
is the non-perturbative vacuum, $\mu,\nu$ are the Dirac spinor
indices. Moreover, $\eta_{\mu}$ is the interpolation field of
nucleons, and for the proton, $
\eta_{\rm{p}}(x)=2[t\eta_1^{\rm{p}}(x)+\eta_2^{\rm{p}}(x)]$, where
two independent terms are given by $
\eta_1^{\rm{p}}(x)=\varepsilon_{abc}[\rm{u}_a^{\rm{T}}\rm{C}\gamma_5\rm{d}_b(x)]\rm{u}_c(x)$
and $
\eta_2^{\rm{p}}(x)=\varepsilon_{abc}[\rm{u}_a^{\rm{T}}\rm{C}\rm{d}_b(x)]\gamma_5\rm{u}_c(x)$,
with C the charge conjugate operator, and $t$ called the Ioffe
parameter whose value is around $-1$\,\cite{Iof81}. In this work,
the value of $t$ is determined via the nucleon mass in
vacuum\,\cite{Cai17x}. In order to obtain the interpolation field
for neutron, one can make the exchange
``$\rm{u}\leftrightarrow\rm{d}$".

In Eq.\,(\ref{SP}),
$\rho_{\mu\nu}=(2\pi)^{-1}\int\d^4xe^{iqx}\langle
0|\eta_{\mu}(x)\overline{\eta}_{\nu}(0)|0\rangle$ and
$\widetilde{\rho}_{\mu\nu}=(2\pi)^{-1}\int\d^4xe^{iqx}\langle
0|\overline{\eta}_{\nu}(0)\eta_{\mu}(x)|0\rangle$ are nucleon
spectral densities. Moreover, Lorentz symmetry and parity invariance
together indicate that the general structure of the spectral density
is $
\rho_{\mu\nu}(q)=\rho_{\rm{s}}(q^2)\delta_{\mu\nu}+\rho_{\rm{q}}(q^2)\slashed{q}_{\mu\nu}
$\,\cite{Coh95}, where $\rho_{\rm{s}}$ and $\rho_{\rm{q}}$ are two
scalar functions of $q^2$. Correspondingly, we have
\begin{equation}\label{Piab}
\Pi_{\mu\nu}(q)=\Pi_{\rm{s}}(q^2)\delta_{\mu\nu}+\Pi_{\rm{q}}(q^2)\slashed{q}_{\mu\nu}
,\end{equation} where the coefficients are\,\cite{Coh95}
\begin{equation}\label{Poly_for}
\Pi_j(q^2)=\int_0^{\infty}\d
s\frac{\rho_j(s)}{s-q^2}+\rm{polynomials},~~j=\rm{s,~q},
\end{equation}
with $s$ the threshold parameter ($\sim M^2$ for a
nucleon). For example, the simplest phenomenological nucleon spectral densities take the
form $\rho_{\rm{s}}^{\rm{phen}}(s)=M\delta(s-M^2)$ and $\rho_{\rm{q}}^{\rm{phen}}(s)
=\delta(s-M^2)$, corresponding to $\Pi(q)=-(\slashed{q}+M)/(q^2-M^2+i0^+)$, which is the
standard nucleon propagator in vacuum, i.e., the two-point nucleon-nucleon correlation
function.

For two operators $A$ and $B$, the OPE gives $
\rm{T}A(x)B(0)=\sum_nC_n^{AB}(x,\mu)\mathcal{O}_n(0,\mu)$ as
$x\to0$, where $C_n^{AB}$'s are the Wilson's coefficients, which can
be obtained by standard perturbative methods\,\cite{Wil69}, and
$\mu$ is the renormalization energy scale. In the momentum space, we
thus have $ \Pi_j(Q^2)=\sum_nC_n^j(Q^2)\langle\mathcal{O}_n\rangle$,
where $Q^2=-q^2$, and $\langle\mathcal{O}_n\rangle$'s are different
types of quark/gluon condensates\,\cite{Coh95}. We note that the OPE
is only meaningful in the deep space-like region.

Furthermore, for any function of momentum transfer, the Borel
transformation $ \mathcal{B}[f(Q^2)]\equiv
\widehat{f}(\mathscr{M}^2)$ is defined by\,\cite{Coh95}
\begin{equation}
\widehat{f}(\mathscr{M}^2)\equiv\lim_{\substack{Q^2,n\to\infty\\
Q^2/n=\mathscr{M}^2}}\frac{(Q^2)^{n+1}}{n!}\left(-\frac{\d}{\d
Q^2}\right)^nf(Q^2),\end{equation} where $\mathscr{M}\sim M$ is the
Borel mass\,\cite{Iof81}. Under the Borel transformation, the
correlation function Eq.\,(\ref{Poly_for}) becomes
\begin{equation}\label{BT}
\widehat{\Pi}_j(\mathscr{M}^2)=\int_0^{\infty}\d
se^{-s/\mathscr{M}^2}\rho_j(s),~~j=\rm{s,~q},\end{equation} where
polynomials in Eq.\,(\ref{Poly_for}) disappear.

After making the Borel transformation on the correlation functions
both from the phenomenological side (i.e., $\Pi^{\rm{phen}}$, which
encapsulates information of the spectral densities) and from the
OPE $(\Pi^{\rm{OPE}})$ under some assumptions\,\cite{Coh95}, we
obtain the QCDSR equations apparently relating the nucleon
self-energies and correspondingly the $E_{\rm{n}}(\rho)$ via
Eq.\,(\ref{EOSPNM}) on the phenomenological side, and the
quark/gluon condensates on the OPE
side\,\cite{Cai17x,Coh91,Coh95}. Physically,
the correlation functions from OPE are not the same as those from the
phenomenological side, and they may even be very different from each
other. The basic assumption of QCDSR is that in some range of $q^2$,
these different correlation functions are the same, in the sense
that the physical quantities are insensitive to the Borel mass
$\mathscr{M}$ introduced\,\cite{Coh95}. This range of $\mathscr{M}$ is often called
the QCDSR window\,\cite{Iof81,Coh95}.

It should be pointed out that QCDSR will become a little difficult as
density/momentum increases for neutron matter problem. The spectral
densities in nuclear medium are very complicated owing to the
complicated medium effects (such as excitations and correlations),
and only at low densities/momenta there exists a narrow resonance
state (the $\delta$-peak) corresponding to the nucleon degree of freedom
($\rho_{\rm{s}}\sim M\delta(s-M^2)+\cdots$ and
$\rho_{\rm{q}}\sim\delta(s-M^2)+\cdots$). As density/momentum
increases, continuum excitations will eventually emerge and these
high density/momentum states will have increasing importance at high
densities/momenta. While on the other hand, in QCDSR, contributions
from these poorly-known complicated high order states are suppressed by Borel
transformation of the correlation functions (characterized by the
factor $e^{-s/\mathscr{M}^2}$), and they can be even removed (as the
polynomials in Eq.\,(\ref{Poly_for})).
As a rough estimate on the density region above which the QCDSR for nucleonic matter becomes broken down,
we consider the formation of the $\Delta$ resonance as an excited state in dense nucleonic matter.
As shown in Ref.\,\cite{Cai15xx}, the formation density of the first charged state of $\Delta(1232)$
could be smaller than 2$\rho_0$, even to be around the saturation density. Thus it is
conservative to expect that the QCDSR for nucleonic matter should not be applied at
densities around or above 2$\rho_0$.
However, a comprehensive analysis of the applicable region of the conventional
QCDSR for nucleonic matter  deserve more further work.

At finite densities, a new term proportional to the nucleon velocity, i.e., $\Pi_{\rm{u}}(q^2,qu)\slashed{u}_{\mu\nu}$ with
$qu=q_{\mu}u^{\mu}$\,\cite{Coh95},
should be added to Eq.\,(\ref{Piab}). Similarly, the correlation
functions constructed from quark/gluon condensates are then given by
\begin{equation}
\Pi_j(q^2,qu)=\sum_nC_n^j(q^2,qu)\langle\mathcal{O}_n\rangle_{\rho},
\end{equation}
where $ \langle\mathcal{O}_n\rangle_{\rho}$ are the condensates
at finite densities\,\cite{Jin93,Coh95}.

In this work, the quark/gluon condensates at finite densities up to mass dimension-6
are included in the QCDSR equations, i.e., $\langle\overline{q}q\rangle$,
$\left\langle({\alpha_{\rm{s}}}/{\pi})G^2\right\rangle$,
$\langle
g_{\rm{s}}\overline{q}\sigma\mathcal{G}q\rangle$, $\langle
g_{\rm{s}}{q}^{\dag}\sigma\mathcal{G}q\rangle$, $\langle\overline{q}\Gamma_1q\overline{q}\Gamma_2q\rangle
$ and $
\langle\overline{q}\Gamma_1\lambda^Aq\overline{q}\Gamma_2\lambda^Aq\rangle$, see Refs.\,\cite{Jeo13,Cai17x} for
more details. For the very relevance for the discussion in this
paper, we write down the expression for the quark
condensates, i.e.,
\begin{equation}\label{chiral_cond}
\langle\overline{q}q\rangle_{\rho,\delta}\approx\langle\overline{q}q\rangle_{\rm{vac}}
+\frac{\sigma_{\rm{N}}}{2m_{\rm{q}}}\left(1\mp\xi\delta\right)\rho
+\Phi(1\mp g\delta)\rho^2 ,
\end{equation}
where ``$-$" (``+") is for the u (d) quark,
$\delta=(\rho_{\rm{n}}-\rho_{\rm{p}})/(\rho_{\rm{n}}+\rho_{\rm{p}})$
is the isospin asymmetry of neutrons and protons in asymmetric
nucleonic matter (ANM) with $\rho_{\rm{n/p}}$ the neutron/proton
density. The corresponding condensate in vacuum takes $
\langle\overline{q}q
\rangle_{\rm{vac}}\approx-(252\,\rm{MeV})^3$\,\cite{Coh95}.
Moreover, $\xi\approx0.1$ characterizing the density dependence of
the condensates for different quarks is fixed by the mass relation
of the baryon octet\,\cite{Jeo13},
$m_{\rm{q}}\equiv (m_{\rm{u}}+m_{\rm{d}})/2 \approx3.5\,\rm{MeV}$
is the average mass of two light quarks, and $\sigma_{\rm{N}}\equiv
m_{\rm{q}}{\d M}/{\d m_{\rm{q}}}\approx45\,\rm{MeV}$ is the
pion-nucleon sigma term\,\cite{Gas91}.

The motivation of including the last term ``$\Phi(1\mp
g\delta)\rho^2$" in Eq.\,(\ref{chiral_cond}) is as follows: As the
density increases, the linear density approximation for the chiral
condensates becomes worse eventually, and
high order terms in density should be included in the
$\langle \overline{q}q\rangle_{\rho,\delta}$. However, the density
dependence of the chiral condensates is extremely complicated, and
there is no general power counting scheme to incorporate these
high density terms. Besides the $\rho^2$ term we adopted here, for instance,
based on the chiral effective theories\,\cite{Kai09,Kru13a}, a term
proportional to $\rho^{5/3}$ was found in the perturbative expansion of
$\langle\overline{q}q\rangle_{\rho,\delta}$ in
$\rho$. On the other hand, using the chiral Ward identity\,\cite{God13}, a $\rho^{4/3}$ term
was found in the density expansion in the chiral condensates.
In our work, {including the higher-order $\rho^2$ term} is mainly for the improvement
{of describing the empirical} EOS of PNM {around and above saturation density,
for which we use the celebrated Akmal--Pandharipande--Ravenhall (APR) EOS\,\cite{APR}.}
In this sense, the $\Phi$-term we adopted here {is} an effective
correction to the chiral condensates beyond the linear {leading-}order.
Two aspects related to the $\Phi$-term should be pointed out: 1).
Without the {higher-order $\rho^2$} term, the EOS of PNM {around and above saturation density} can not
be adjusted to be consistent with that of APR EOS, i.e., there exists
systematic discrepancy between the QCDSR EOS and the APR EOS around and above saturation density;
2). Using an effective correction with a different power in density, e.g., a $\rho^{5/3}$ {term},
the conclusion does not change, i.e., the EOS
of PNM around and above saturation density can still be adjusted to fit the APR EOS,
and the sign of the coefficients $\Phi$ and $g$ does not change,
and this will be seen from Fig.\,\ref{fig_Fig1} in the following.
Moreover, the physical origin of the high density term in the chiral
condensates is an interesting issue, and one of the possibilities is
the three-body force. For instance, in the Skyrme--Hartree--Fock (SHF) model,
a traditional two-body force contributes a term proportional to $\rho$ to
the EOS, and a $\rho^{1+\alpha}$ term emerges once the effective
three-body force is considered\,\cite{Zha16}. Here $\alpha$ is a parameter
characterizing the three-body force.
Exploring the three-body force in the QCDSR\,\cite{Dru17} and
its connection to the high density term in the chiral condensates
will be useful for further applications of the QCDSR in nucleonic matter calculations.
In the following, we abbreviate the
QCDSR using the chiral condensate without the last term in
Eq.\,(\ref{chiral_cond}) in ``naive QCDSR''.

Furthermore, the four-quark condensate used in this work takes the conventional
decomposition structure as
\begin{equation}\label{f4}
\widetilde{\langle\overline{q}q\rangle_{\rho,\delta}^2}=(1-f)\langle\overline{q}q\rangle_{\rm{vac}}^2
+f\langle\overline{q}q\rangle_{\rho,\delta}^2,\end{equation} where
$f$ is an effective parameter introduced in Refs.\,\cite{Jin93,Jeo13,Coh95}.
Besides the above input on the chiral/conventional four-quark condensates, the other condensates are adopted as the same as those in
Refs.\,\cite{Jin93,Coh95,Jeo13}.
Effects of twist-four four-quark condensates\,\cite{Jeo13} on the $E_{\rm{n}}(\rho)$
are not considered and will be discussed at the end of the next section.
Finally, in carrying out the QCDSR calculations, we fix the central
value of the $E_{\rm{n}}(\rho)$ at a very low density
$\rho_{\rm{vl}}=0.02\,\rm{fm}^{-3}$ to be consistent with the
prediction by the chiral perturbative theories (ChPT)\,\cite{Tew13,Kru13},
i.e., $E_{\rm{n}}(\rho_{\rm{vl}})=4.2\,\rm{MeV}$,
the central value of the symmetry energy $E_{\rm{sym}}(\rho)$ at a critical density
$\rho_{\rm{c}}=0.11\,\rm{fm}^{-3}$ to be $E_{\rm{sym}}(\rho_{\rm{c}}) = 26.65$ MeV\,\cite{ZhangZ13},
and fit the EOS of PNM to the APR EOS, via varying $\Phi$, $g$ and $f$.
{We note that} the parameter $f$ is {essentially} determined by
$E_{\rm{n}}(\rho_{\rm{vl}})$, and the overall fitting of the EOS of PNM
to the APR EOS and the symmetry energy at $\rho_{\rm{c}}$
determines the other two parameters $\Phi$ and $g$.

\section{EOS of PNM from QCDSR}\label{Sec3}

\begin{figure*}[tbh!]
\centering
  \includegraphics[width=12cm]{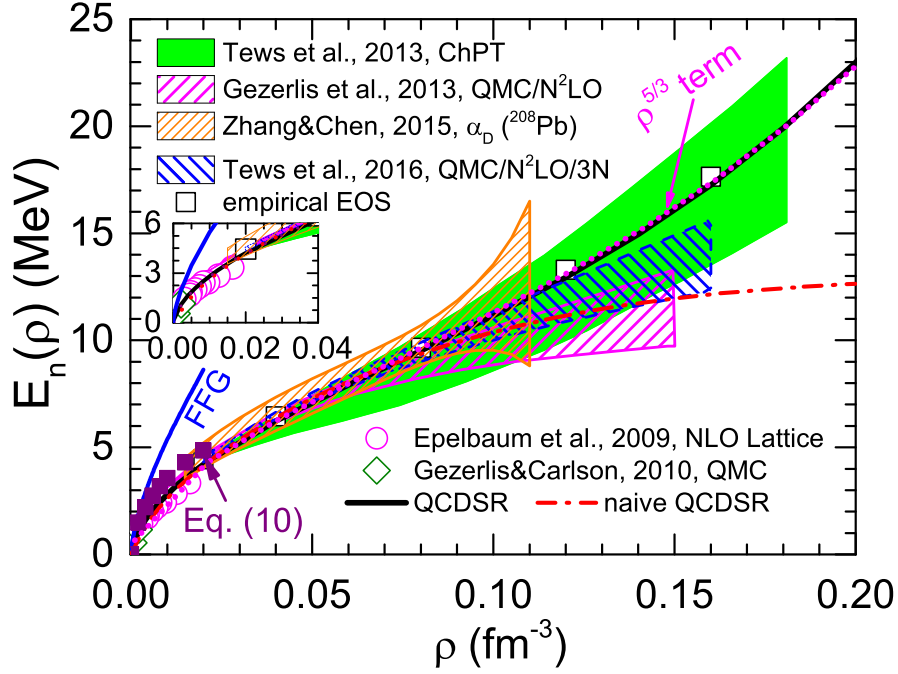}
  \caption{(Color Online). EOS of PNM obtained by QCDSR and by the naive QCDSR. Results from other approaches are also shown for comparison (see the
  text for details).}
  \label{fig_Fig1}
\end{figure*}

In Fig.\,\ref{fig_Fig1}, we show the predictions on the
$E_{\rm{n}}(\rho)$ by QCDSR with
$\langle\overline{q}q\rangle_{\rm{vac}} = -(252\,\rm{MeV})^3$,
$\xi = 0.1$, $m_{\rm{q}} = 3.5\,\rm{MeV}$, $\sigma_{\rm{N}} = 45\,\rm{MeV}$,
$\Phi'\equiv\Phi\times\langle\overline{q}q\rangle_{\rm{vac}}=3.45$,
$g=-0.64$ and $f=0.43$.
In {the} case of the naive QCDSR, we fix $E_{\rm{n}}(\rho_{\rm{vl}})=4.2\,\rm{MeV}$ via varying
the $f$ parameter, {and find} $f=0.50$.
Also included in Fig.\,\ref{fig_Fig1} are the results from
ChPT\,\cite{Tew13,Kru13} (green band), quantum Monte Carlo (QMC)
simulations {combined} with chiral force to next-to-next-to-leading
order (N$^2$LO) with\,\cite{Tew16} (blue band) and
without\,\cite{Gez13} (magenta band) leading-order chiral
three-nucleon interactions forces, next-to-leading order (NLO)
lattice calculation\,\cite{Epe09a} (magenta circle), {and} QMC simulations
for PNM at very low densities\,\cite{Gez10} (green diamond). The
result from analyzing experimental data on the electric dipole polarizability $\alpha_{\text D}$ in
$^{208}$Pb\,\cite{Zha15} is also shown for comparison.
Based on the obtained $\Phi$ and $g$,
we can estimate the density below which the $\Phi$-term
has minor contribution to the quark condensates.
This density can be estimated from $|\Phi(1-g)\rho^2|\ll|(\sigma_{\rm{N}}/2m_{\rm{q}})(1-\xi)\rho|$,
i.e., the last term in Eq.\,(\ref{chiral_cond}) is significantly
less than the second term in Eq.\,(\ref{chiral_cond}),
and we thus obtain
$\rho\ll\rho_{\rm{es}}\approx2.13\,\rm{fm}^{-3}$.
Therefore, the effects of $\Phi$ and $g$ on the $E_{\rm{n}}(\rho)$ are
trivial at subsaturation densities, e.g., when one artificially
takes $\Phi'=0$ and keeping $f$ fixed, the $E_{\rm{n}}(\rho_{\rm{vl}})$ $(E_{\rm{n}}(0.1\,\rm{fm}^{-3})$ changes
from 4.20\,MeV to 4.22\,MeV (from 11.15\,MeV to 10.01\,MeV).
It is thus reasonable to expect that effects of $\Phi$ and
$g$ on the $E_{\rm{n}}(\rho)$ at low densities $\lesssim0.1\,\rm{fm}^{-3}$ are small.
However, as the density increases, there is no guarantee that the $\Phi$-term
still has small effects on the EOS of PNM since the $E_{\rm{n}}(\rho)$ is
obtained by integrating over the density (see Eq.\,(\ref{EOSPNM})).

The inset in Fig.\,\ref{fig_Fig1} shows the EOS of PNM at very low
densities {where the results are almost the same for} the QCDSR and {the} naive QCDSR.
Actually, after neglecting the contributions from dimension-4 and higher order terms,
we can obtain an analytical approximation for EOS of PNM as\,\cite{Cai17x},
\begin{equation}\label{LPNM}
E_{\rm{n}}(\rho)\approx
E_{\rm{n}}^{\rm{FFG}}(\rho)-\frac{\rho}{2}\frac{M}{\langle\overline{q}q\rangle_{\rm{vac}}}
\left(5-\frac{\sigma_{\rm{N}}}{2m_{\rm{q}}}+\frac{\xi\sigma_{\rm{N}}}{2m_{\rm{q}}}\right),
\end{equation}
where $E_{\rm{n}}^{\rm{FFG}}(\rho)=3k_{\rm{F}}^{\rm{n},2}/10M\sim\rho^{2/3}$ is
the free Fermi gas (FFG) prediction. Eq.\,(\ref{LPNM}) clearly
demonstrates how the chiral condensate goes into play in the EOS
of PNM, i.e., the second term characterized by several constants
($\xi,\sigma_{\rm{N}},m_{\rm{q}}$ and
$\langle\overline{q}q\rangle_{\rm{vac}}$) is negative, leading to a
reduction on the $E_{\rm{n}}(\rho)$ compared to the FFG prediction.
In Fig.\,\ref{fig_Fig1}, we also plot the results obtained from Eq.\,(\ref{LPNM})
at densities $\lesssim0.02\,\rm{fm}^{-3}$ (violet solid square).
One can see that the approximation Eq.\,(\ref{LPNM}) can already produce
reasonably the $E_{\rm{n}}(\rho)$ at low densities.
Furthermore, it is seen from Fig.\,\ref{fig_Fig1} that the prediction on the
$E_{\rm{n}}(\rho)$ from QCDSR is consistent with several QMC
simulations and lattice computation,
showing QCDSR is a reliable approach in the study of PNM, especially at
lower densities, where the naive QCDSR is good enough.

Another feature of Fig.\,\ref{fig_Fig1} is that compared with the APR EOS,
the prediction on the EOS of PNM in the naive QCDSR is well-behaved for
$\rho\lesssim0.1\,\rm{fm}^{-3}$. However,
as the density increases, the discrepancy between the overall shape
of $E_{\rm{n}}(\rho)$ predicted by the naive QCDSR and by the APR
becomes large and this can not be improved by adjusting the parameter $f$ in the naive QCDSR,
indicating that the leading-order linear density approximation for
the chiral condensates dose not work well enough {and the higher order density terms
in the chiral condensates are needed} for PNM calculations
in the density region of $\rho\gtrsim0.1\,\rm{fm}^{-3}$. Once we consider the term $\Phi
g\rho^2$ in Eq.\,(\ref{chiral_cond}) for PNM, and recalculate the EOS of
PNM, {we find that compared with the case of the naive QCDSR,} the obtained prediction
can be largely improved to fit the APR EOS. For example, the EOS of
PNM at $0.12\,\rm{fm}^{-3}$ is now found to be 12.9\,MeV, which is very close to
the APR prediction 13.3\,MeV. This feature suggests that the QCDSR with
effective higher order density terms in quark condensates can be
used to study the EOS of dense nucleonic matter at higher densities.
It is necessary to point out that using a different high density term
in Eq.\,(\ref{chiral_cond}) and re-fix the parameters $f$, $\Phi$ and $g$
by the same fitting scheme, the density behavior of the $E_{\rm{n}}(\rho)$
is almost unchanged. For instance, when adopting a $\rho^{5/3}$ term, i.e.,
$\Phi(1\mp g\delta)\rho^{5/3}$, we then obtain $f\approx0.46$,
$\Phi'\equiv \Phi\times\langle\overline{q}q\rangle_{\rm{vac}}^{2/3}\approx1.61$
and $g\approx-0.34$, and the corresponding $E_{\rm{n}}(\rho)$ is shown in
Fig.\,\ref{fig_Fig1} by the magenta dot line.
It is clearly seen that using a different high density term in the
chiral condensates will not change our conclusions on the EOS of PNM.

Furthermore, it should be noted that once the twist-four four-quark
condensates~\cite{Jeo13} are included in the QCDSR equations and
the $E_{\rm{n}}(\rho)$ is still fixed at $0.02\,\rm{fm}^{-3}$ and made to be
consistent with the APR EOS as much as possible, we find that the EOS of PNM
at densities $\lesssim0.12\,\rm{fm}^{-3}$ is essentially the same as the one
without these condensates.
And at nuclear saturation density $\rho_0 = 0.16\,\rm{fm}^{-3}$, the $E_{\rm{n}}(\rho_0)$ changes from about 17.1\,MeV to 15.9\,MeV.
As the high-twist operators have some impacts on several processes in hadronic
physics\,\cite{Gre07}, the exact knowledge on density dependence of {the EOS of PNM}
may provide a novel tool to study them. Since including the twist-four four-quark
condensates does not affect our present conclusions, we will not discuss effects of
these terms again in the following sections and leave the details to be reported
elsewhere\,\cite{Cai17x}.

Finally, we would like to briefly discuss the properties of the EOS of symmetric
nuclear matter (SNM) obtained in the QCDSR.
Although our main point on the above fitting scheme is the EOS of PNM at
densities of $\rho \lesssim\rho_0$ and the symmetry energy at $\rho_{\rm{c}}$ with
the inclusion of an effective correction in $\rho^2$, the predictions on the
saturation properties of the SNM are significantly improved from
$(\rho_0,E_0(\rho_0))\approx(0.6\,\rm{fm}^{-3},-99\,\rm{MeV})$ in the naive
QCDSR to $(0.2\,\rm{fm}^{-3},-26\,\rm{MeV})$ {in the QCDSR}. It suggests from
another viewpoint that the effective $\Phi$-term in Eq.\,(\ref{chiral_cond})
is important, implying the breakdown of the chiral condensates at linear
order at densities even smaller than the saturation density.
In fact, it is a challenging problem on how to improve the saturation properties
of the SNM in the microscopic theories (see, e.g., Ref.\,\cite{Dri17}).
Improvement on the saturation properties of the SNM in the QCDSR is important,
and this is beyond the main motivation of the present work.

\section{Chiral Condensates}\label{Sec4}

\begin{figure}[tbh!] \centering
\includegraphics[width=8.5cm]{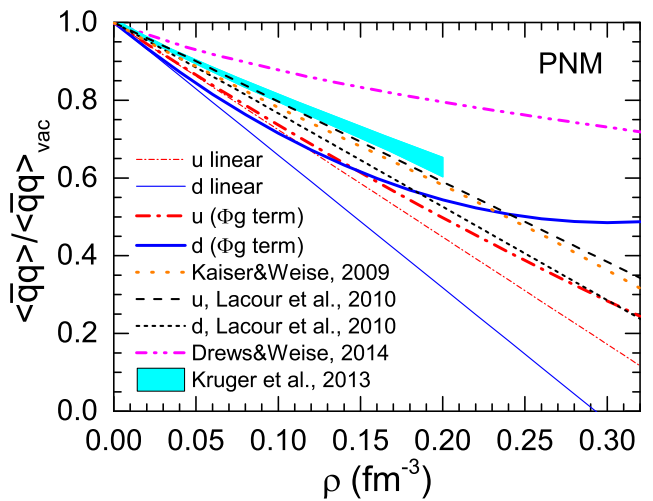}
\caption{(Color Online). Density dependence of quark condensates in PNM from
QCDSR. Also shown are the results from ChPT\,\cite{Kru13a,Lac10,Kai09} and FRG
approach\,\cite{Dre14}.}
  \label{fig_Fig2}
\end{figure}

In Fig.\,\ref{fig_Fig2}, we show the density dependence of the
quark condensates from QCDSR as well as {the corresponding} predictions from
ChPT\,\cite{Kru13a,Lac10,Kai09} {and} the functional renormalization
group (FRG) approach\,\cite{Dre14}.
At low densities, the chiral
condensate is dominated by the linear density term. Specifically, we have
$(\langle\overline{\rm{u}}\rm{u}\rangle_{\rho}
-\langle\overline{\rm{d}}\rm{d}\rangle_{\rho})/\langle\overline{q}q\rangle_{\rm{vac}}\approx
-\rho\sigma_{\rm{N}}\xi/
m_{\rm{q}}\langle\overline{q}q\rangle_{\rm{vac}}>0$ at low
densities, since $\langle\overline{q}q\rangle_{\rm{vac}}$ is
negative. As density increases, the $\Phi$ term in
Eq.\,(\ref{chiral_cond}) begins to dominate and even to flip the
relative relation {of the magnitude between $\langle\overline{\rm{u}}\rm{u}\rangle_{\rho}$ and
$\langle\overline{\rm{d}}\rm{d}\rangle_{\rho}$, leading to}
$\langle\overline{\rm{u}}\rm{u}\rangle_{\rho}/\langle\overline{q}q\rangle_{\rm{vac}}<
\langle\overline{\rm{d}}\rm{d}\rangle_{\rho}/\langle\overline{q}q\rangle_{\rm{vac}}$
{when the density $\rho$ is larger than about} $0.15\,\rm{fm}^{-3}$. For example,
$\langle\overline{\rm{d}}\rm{d}\rangle_{\rho_0}/\langle\overline{q}q\rangle_{\rm{vac}}\,(\langle\overline{\rm{u}}\rm{u}\rangle_{\rho_0}/\langle\overline{q}q\rangle_{\rm{vac}})$
in PNM changes from $0.45$ ($0.56$) in the linear density approximation
to $0.60$ ($0.59$) {with the inclusion of the $\Phi$ term in Eq.\,(\ref{chiral_cond})},
leading to an enhancement {of} about 33\% (5\%). It is interesting to
point out that the flip is a direct consequence of the
{inclusion of the higher order $\Phi$ term in Eq.\,(\ref{chiral_cond})}.

Furthermore, it is interesting to see that the high order $\Phi$ term in
Eq.\,(\ref{chiral_cond}) tends to stabilize the chiral condensate both for u and d
quarks at higher densities, while the leading-order linear density
approximation Eq.\,(\ref{chiral_cond}) leads to chiral symmetry restoration
at a density of about $2\rho_0$.
This hindrance of the chiral symmetry restoration due to the high order density
terms in quark condensates has important implications on the physical degrees
of freedom in the core of neutron stars where the matter is very close to PNM.
This feature is consistent with the recent analysis on the same issue using
the FRG method~\cite{Dre14}.

At this point, we would like to discuss the role played by the $\sigma_{\rm{N}}$.
In our calculations above, the value of $\sigma_{\rm{N}}$ is fixed at 45\,MeV.
The physical value of $\sigma_{\rm{N}}$ still has a sizable uncertainty.
With a different $\sigma_{\rm{N}}$, however, we need to readjust the values of the parameters
$f$, $\Phi$ and $g$ based on the fitting scheme we adopted above, i.e., fixing the physical value
of the EOS of PNM at $\rho_{\rm{vl}}$ and the symmetry energy at $\rho_{\rm{c}}$, and
meanwhile fitting the $E_{\rm{n}}(\rho)$ to the APR EOS.
Consequently, in this way, the $\sigma_{\rm{N}}$ has very little {influence}
on the EOS of PNM.
Different {values of} $\sigma_{\rm{N}}$ {will} lead to different values of
$\Phi$ and $g$, but the density dependence of the chiral condensates will change only
quantitatively, instead of qualitatively since the $\sigma_{\rm{N}}$ term (linear order)
is a perturbation to the vacuum chiral condensates.
Similarly, the $\Phi$-term is a perturbation to the linear term.
Besides the quantities involved in the fitting scheme, the $\sigma_{\rm{N}}$
{will also have influence} on {some other} quantities such as the density dependence
of the nucleon effective mass, which will be explored in detail elsewhere\,\cite{Cai17x}.
Finally, {it should be mentioned that} the study on the $\sigma_{\rm{N}}$ itself is an important issue, and
it will help improving our understanding on the relevant aspects of the strong interaction.

\section{Summary}\label{Sec5}

We have studied the EOS of PNM $E_{\rm{n}}(\rho)$ within the framework of QCDSR
by effectively taking into account the higher-order density effects in the
quark condensates. Firstly, the $E_{\rm{n}}(\rho)$ thus obtained is found to be
consistent with the predictions from current advanced microscopic many-body theories.
Our results have indicated that although the higher-order density
terms in quark condensates play minor role for EOS of PNM at subsaturation densities
($\rho \lesssim0.1$ fm$^-3$), they play an important role in describing the EOS of
PNM in the density region around and above nuclear saturation density.

Secondly, our results have demonstrated that the higher-order density terms in
quark condensates tends to stabilize the u/d chiral condensates at higher
densities, which is consistent with the predictions from other
advanced microscopic many-body calculations. This feature has important implications
on the QCD phase diagram under extreme conditions of low temperatures, large isospin
and large baryon chemical potentials, which is essential for understanding the
physical degrees of freedom in the core of neutron stars.

\section*{Acknowledgments}
This work was supported in part by the National Natural Science
Foundation of China under Grant No. 11625521, the Major State Basic Research
Development Program (973 Program) in China under Contract No.
2015CB856904, the Program for Professor of Special Appointment (Eastern
Scholar) at Shanghai Institutions of Higher Learning, Key Laboratory
for Particle Physics, Astrophysics and Cosmology, Ministry of
Education, China, and the Science and Technology Commission of
Shanghai Municipality (11DZ2260700).

\end{document}